\documentclass[twocolumn,english]{revtex4-1}
\usepackage[T1]{fontenc}
\usepackage[latin9]{inputenc}
\usepackage{float}
\usepackage{amsmath}
\usepackage{graphicx}
\usepackage{esint}

\makeatletter

\providecommand{\tabularnewline}{\\}

\usepackage{babel}

\usepackage{babel}

\usepackage{babel}

\makeatother

\usepackage{babel}
\begin{document}

\title{Role of local response in manipulating the elastic properties of
disordered solids by bond removal}

\author{Daniel Hexner}
\email{danielhe2@uchicago.edu}

\affiliation{The James Franck Institute and Department of Physics, The University
of Chicago, Chicago, IL 60637, USA and Department of Physics and Astronomy,
The University of Pennsylvania, Philadelphia, PA, 19104, USA}

\author{Andrea J. Liu}

\affiliation{Department of Physics and Astronomy, The University of Pennsylvania,
Philadelphia, PA, 19104, USA}

\author{Sidney R. Nagel}

\affiliation{The James Franck and Enrico Fermi Institutes and The Department of
Physics, The University of Chicago, Chicago, IL 60637, USA}
\begin{abstract}
We explore the range over which the elasticity of disordered spring
networks can be manipulated by the removal of selected bonds. By taking
into account the local response of a bond, we demonstrate that the
effectiveness of pruning can be improved so that auxetic (i.e., negative
Poisson's ratio) materials can be designed without the formation of
cracks even while maintaining the global isotropy of the network.
The bulk, $B$, and shear, $G$, moduli scale with the number of bonds
removed and we estimate the exponents characterizing these power laws.
We also find that there are spatial correlation lengths in the change
of $B$ and $G$ upon removing different bonds that diverge as the
network approaches the isostatic limit where the excess coordination
number $\Delta Z\rightarrow0$. 
\end{abstract}
\maketitle

\section{Introduction}

Manipulating the elastic properties of solids is an important problem
with broad applications~\cite{Nagel2017,DesignerPerspective}. The
most common approach in designing mechanical metamaterials is based
on a periodically repeating unit cell that is carefully constructed
to yield a given elastic property or function. Recently, a novel design
principle has been introduced based on ``pruning'' disordered spring
networks~\cite{tuning_by_pruning}. This exploits the broad distribution
of how different bonds contribute to the elastic moduli in such systems
\textendash{} by selectively removing a bond that contributes more
to one modulus than to another, one can prune a system to achieve
desired elastic properties. Disorder provides two clear advantages
over periodic lattices: 1) disordered systems are isotropic on large
scales; 2) disorder allows materials to be designed with inhomogeneous
and even local responses~\cite{Rocks2017,Yan2017,Flechsig2017,Yan2017a}.

To demonstrate the potential flexibility that pruning provides, consider
the effect on an elastic modulus of the removal of a single bond in
a spring network with $N_{b}$ bonds. The modulus could characterize
the cost of a global deformation, such as compression, or a local
deformation, such as the pinching together of two nodes. If the system
is periodic, removing a bond in one unit cell will result in the same
change in the elastic moduli as the removing the equivalent bond in
any other unit cell. However, for a disordered network, removing different
bonds leads to different responses. Naively, the removal of the first
bond results in $N_{b}$ different possible responses while removing
$N_{r}$ bonds leads to $N_{b}!/N_{r}!\left(N_{b}-N_{r}\right)!$
which even for small $N_{r}$ can be an enormous number.

To realize this large range of possible designs it is important for
the system to obey two properties. First, removing a bond, $i$, affects
different moduli differently in an uncorrelated way. Indeed this was
shown to be the case for bulk deformations: removal of bond $i$ changes
the bulk modulus by $\Delta B_{i}$ and the shear modulus by $\Delta G_{i}$,
where $\Delta B_{i}$ and $\Delta G_{i}$ have nearly vanishing correlations~\cite{Hexner2017}.
Second, the change in moduli upon the removal of a bond must have
a broad range. This is also the case for bulk deformations, where
$\Delta B_{i}$ and $\Delta G_{i}$ at small values scale as a power-law~\cite{Hexner2017}.

The systems we design are based on disordered networks derived from
jammed packings~\cite{Ohern,Review}. Soft repulsive spherical particles
are placed randomly in space and the energy is minimized to attain
force balance. The centers of the spheres are then connected by springs
to form a network and the equilibrium spring length is set to the
equilibrium distance between nodes, thus removing all stresses. For
simplicity, all the spring constants, $k$, are chosen to be equal.
We characterize ensembles of such networks by the coordination number
per node, $Z$, and the excess coordination number, $\Delta Z=Z-Z_{c}$,
where $Z_{c}$ is the critical value of $Z$ at which rigidity is
lost: in an infinite system, $Z_{c}=2d$~\cite{Durian1995,Goodrichfinitesize,angle_avg}
.

If a bond length, $r_{i}$, between two nodes is different from the
equilibrium length, $r_{i}^{0}$, there is an energy cost of $\frac{1}{2}k\delta^{2}r_{i}$
and a tension $\tau_{i}=k\delta r_{i}$ where $\delta r_{i}\equiv r_{i}-r_{i}^{0}$.
Since the networks are initially unstressed, compressing the system
results in an energy $\frac{1}{2}B\epsilon_{B}^{2}$, where $B$ is
the bulk modulus and $\epsilon_{B}$ is the compression strain; similarly
the energy cost of a shear is $\frac{1}{2}G\epsilon_{G}^{2}$ where
$G$ is the shear modulus and the $\epsilon_{G}$ is the shear strain.
Since the energy is additive, $U=\sum_{i}\frac{1}{2}k\delta^{2}r_{i}$
this allows us to decompose the bulk and shear modulus into a sum
over their single bond contributions: $B=\sum_{i}B_{i}$ and $G=\sum_{i}G_{i}$
where $B_{i}$ and $G_{i}$ are the contributions to $B$ and $G$
respectively of bond $i$~\cite{tuning_by_pruning}.

In an isotropic system, the Poisson's ratio, $\nu$, is a monotonically
decreasing function of $G/B$. In the method introduced by Ref.~\cite{tuning_by_pruning},
bonds are pruned that target either $G$ or $B$ to yield the desired
value of $\nu$. To attain, for example, a large $G/B$, two strategies
may be envisioned. Bonds that contribute a large amount to the bulk
modulus (large values of $B_{i}$) can be successively removed; due
to the relatively weak correlations between $B_{i}$ and $G_{i}$,
this results in only a moderate decrease in $G$ but a large drop
in $B$. Alternatively, bonds that contribute little to $G$ (small
values of $G_{i}$) can be removed so that $G$ does not change appreciably
but again due to the weak correlations, $B$ decreases more steeply.
The first approach has the undesired effect of creating cracks~\cite{TunableFailure}
since the removed bonds carry a lot of stress under the deformation,
which is distributed to its neighbors when removed.

We recently pointed out~\cite{Hexner2017} that there is a difference
between the contribution of a given bond $i$ to the modulus, $M_{i}$,
and the \emph{change} of the modulus if bond $i$ is removed, $\Delta M_{i}$.
Clearly, the evolution of the $M$ under pruning depends on the latter
quantity. Ref.~\cite{tuning_by_pruning} uses $B_{i}$ and $G_{i}$
as proxies for predicting $\Delta B_{i}$ and $\Delta G_{i}$ in order
to tune the values of $B$ and $G$. This approach was quite successful\textendash it
was found that a large $G/B$ can be attained by removing the bonds
with the largest $B_{i}$~\cite{tuning_by_pruning}. Similarly, a
very small $G/B$ can be attained by either removing the bonds with
maximal $G_{i}$ or minimal $B_{i}$. However, removing the minimal
$G_{i}$ strategy fails and both $B$ and $G$ change in a correlated
manner as the bonds are removed. Here, we show that this failure results
from the use of $G_{i}$ as a proxy for $\Delta G_{i}$, and that
when minimal $\Delta G_{i}$ bonds are removed, one can obtain very
large values of $G/B$. Moreover, consideration of $\Delta M_{i}$
instead of $M_{i}$ allows us to estimate theoretically the scaling
exponents of $G/B$ as a function of the number of bonds removed.
We also report correlations in $\Delta B_{i}$ and $\Delta G_{i}$
as a function of the distance between bonds with correlation lengths
$\xi_{\Delta B_{i}}$ and $\xi_{\Delta G_{i}}$ respectively. These
correlation lengths diverge as power laws as the network connectivity
decreases towards the rigidity threshold $\Delta Z=0$.

\section{Tuning $G/B$ }

Here we tune $G/B$ by removing bonds based on $\Delta B_{i}$ and
$\Delta G_{i}$. We explore different pruning strategies that target
bonds that either have a maximal or minimal value of $\Delta B_{i}$
or $\Delta G_{i}$.

The quantities $M_{i}$ and $\Delta M_{i}$ are related via a linear-response
relation~\cite{Hexner2017}: 
\begin{equation}
\Delta M_{i}=M_{i}/S_{i}^{2},\label{eq:DMi}
\end{equation}
where $kS_{i}^{2}$ is local modulus characterizing the cost of a
change in the equilibrium length of bond $i$.

To employ Eq. \ref{eq:DMi} to tune $G/B$, one needs to compute $B_{i}$,
$G_{i}$ and $S_{i}^{2}$ for each bond. Evaluation of $S_{i}^{2}$
for each bond requires $N_{b}$ calculations, however this is done
only once for the initial unpruned network. Thereafter, the evolution
of the spring network is efficiently computed using methods described
in Appendix A. We note that there are $(\frac{d\left(d+1\right)}{2}-1)$
independent shear moduli in $d$ dimensions, which are denoted as
$G^{\left(j\right)}$. Pruning based on their average, denoted by
$G=\frac{1}{\frac{d\left(d+1\right)}{2}-1}\sum_{j}G^{\left(j\right)}$,
allows tuning of $G/B$ in a manner that leaves the system isotropic.
We note that $\Delta G_{i}$ is a linear average over all $\Delta G_{i}^{\left(j\right)}$,
which will be important for determining its distribution.

\begin{figure}[H]
\begin{centering}
\includegraphics[scale=0.4]{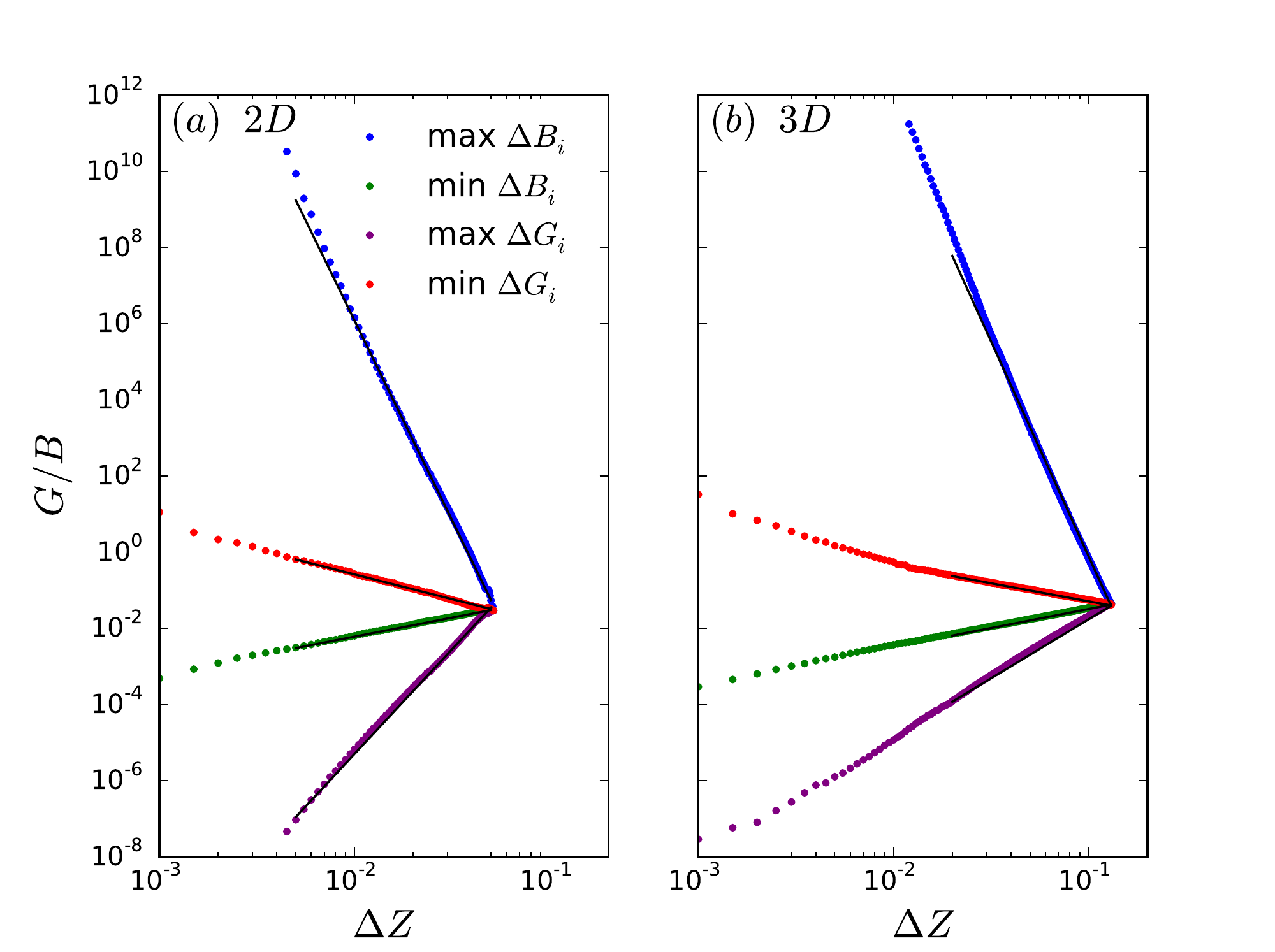} 
\par\end{centering}
\caption{The evolution of $G/B$ as a function of $\Delta Z$ which signifies
the number of bonds removed (a) in two dimensions. (b) in three dimensions
\label{fig:G_B}}
\end{figure}

In Fig.~\ref{fig:G_B} the evolution of $G/B$ is shown for different
pruning strategies in which we remove bonds with: $max\Delta B_{i}$,
$min\Delta B_{i}$, $max\Delta G_{i}$ or $min\Delta G_{i}$. Results
in dimensions $d=2$ and $d=3$ are shown. The number of removed bonds
is characterized by $\Delta Z=\Delta Z_{0}-2N_{r}/N$ where $\Delta Z_{0}$
is the initial excess coordination number and $N_{r}$ are the number
of bonds removed. The pruning procedures based on $min\Delta B_{i}$
and $max\Delta G_{i}$ result in a very small $G/B$ ratio while those
based on $max\Delta B_{i}$ and $min\Delta G_{i}$ result in a very
large $G/B$. We emphasize that only a few percent of bonds are removed,
yet the change in $G/B$ can be almost ten orders of magnitude.

To characterize the variation of $G/B$ versus $\Delta Z$, we fit
the values of $G/B$ for each pruning strategy to a power-law $\Delta Z^{\mu}$.
As illustrated by the larger exponents shown in Table \ref{tab:expon_table},
pruning strategies based on $\Delta B_{i}$ and $\Delta G_{i}$ are
more effective in changing $G/B$ than those based on $B_{i}$ and
$G_{i}$. Moreover, pruning the $min\Delta G_{i}$ bonds yields auxetic
($\nu<0$) behavior, which had not been achieved using $minG_{i}$.

\begin{table}[H]
\begin{centering}
\begin{tabular}{|c|c|c|c|c|c|c|c|c|}
\hline 
pruning method  & $\mu_{B_{+}}^{2D}$  & $\mu_{B_{-}}^{2D}$  & $\mu_{G_{+}}^{2D}$  & $\mu_{G_{-}}^{2D}$  & $\mu_{B_{+}}^{3D}$  & $\mu_{B_{-}}^{3D}$  & $\mu_{G_{+}}^{3D}$  & $\mu_{G_{-}}^{3D}$\tabularnewline
\hline 
\hline 
$B_{i}$, $G_{i}$  & $-5.36$  & $1.27$  & $3.05$  & -  & $-7.96$  & $1.01$  & $1.82$  & -\tabularnewline
\hline 
$\Delta B_{i}$, $\Delta G_{i}$  & $-10.5$  & $1.0$  & $5.5$  & $-1.3$  & $-11.3$  & $1.0$  & $3.1$  & $-0.95$\tabularnewline
\hline 
\end{tabular}
\par\end{centering}
\caption{A comparison of the exponents, defined by $\frac{G}{B}\propto\Delta Z^{\mu}$,
when bonds are selected based on $\Delta B_{i}$ and $\Delta G_{i}$
values versus when they are selected by their $B_{i}$ and $G_{i}$
values, taken from Ref. \cite{tuning_by_pruning}. The subscript of
$\mu$ designates the targeted modulus while $+$ ($-$ ) marks the
maximal (minimal) values targeted. \label{tab:expon_table}}
\end{table}

Equation~\ref{eq:DMi} explains why the procedure based on removing
the bond with $minG_{i}$ was unsuccessful. Because of the appearance
of $S_{i}^{2}$ in the denominator, a bond with a small $G_{i}$ does
not in general have a small $\Delta G_{i}$ . Reference~\cite{Hexner2017}
shows that $G_{i}\propto S_{i}^{2}$ at small $S_{i}^{2}$. Thus bonds
that seem unimportant and carry little stress may in fact be important;
their removal can vary $G$ significantly.

To highlight the difference between $\Delta G_{i}$ and $G_{i}$ we
now consider their distributions. Ref. \cite{tuning_by_pruning} shows
that the distribution of $G_{i}$ at small values scales as a power-law,
which to a good approximation in three dimensions is given by, $G_{i}^{\approx-0.38}$.
This suggests that there are many bonds which can be removed with
little change to the shear modulus. We now argue that this is not
the case, as will be inferred from the $\Delta G_{i}$ distribution.

To compute the distribution of $\Delta G_{i}$ we employ the analysis
of Ref.~\cite{Hexner2017} which studied the distribution of $\Delta G_{i}^{\left(j\right)}$
for any shear direction. It was shown numerically, and supported by
theoretical arguments that to a good approximation: 
\begin{equation}
P\left(\frac{\Delta G_{i}^{\left(j\right)}}{\left\langle \Delta G_{i}^{\left(j\right)}\right\rangle }=y\right)=\frac{1}{\sqrt{2\pi}}y^{-\frac{1}{2}}e^{-\frac{y}{2}}.\label{eq:dist_DG_i}
\end{equation}
The distribution of $\Delta G_{i}$ is then given by the sum over
$n=\frac{d\left(d+1\right)}{2}-1$ different shear directions, $\Delta G_{i}^{\left(j\right)}$.
Assuming that these are independent, the distribution of $\Delta G_{i}$
is computed in \ref{appendix_sum_rand_var} and found to be a Gamma
distribution:

\begin{equation}
P\left(\frac{\Delta G_{i}}{\left\langle \Delta G_{i}\right\rangle }=y\right)=\left(\frac{n}{2}\right)^{n/2}\frac{1}{\Gamma\left(\frac{n}{2}\right)}y^{\frac{n}{2}-1}e^{-\frac{ny}{2}},\label{eq:P_DG_i}
\end{equation}
where $\Gamma\left(n\right)$ is the Gamma function. The important
observation is that at small values $P\left(\Delta G_{i}\right)\propto\Delta G_{i}^{\frac{d\left(d+1\right)}{4}-\frac{3}{2}}$,
so that in three dimensions $P\left(\Delta G_{i}\right)\propto\Delta G_{i}^{+\frac{3}{2}}$
in contrast to the $P\left(G_{i}\right)\propto G_{i}^{\approx-0.38}$
which has the opposite sign in the exponent. Thus, most bonds thought
to be unimportant based on their $G_{i}$ actually lead to a substantial
decrease in $G$.

\begin{figure}[H]
\includegraphics[scale=0.4]{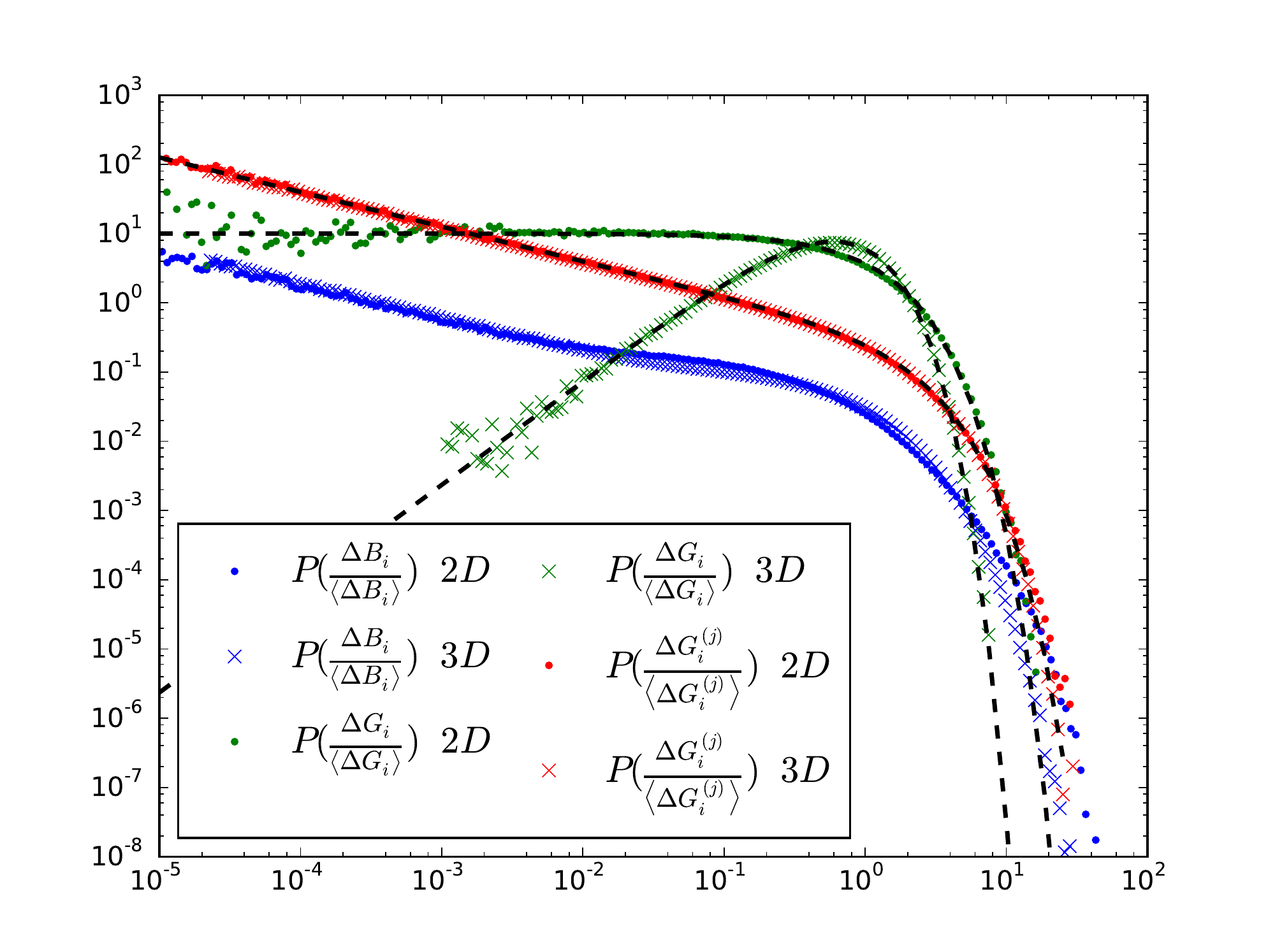}

\caption{The distribution of $\Delta B_{i}$ , $\Delta G_{i}$ and $\Delta G_{i}^{\left(j\right)}$
in both two and three dimensions. The dashed curves overlaying $P\left(\Delta G_{i}^{\left(j\right)}\right)$
is the prediction from Ref. \cite{Hexner2017} : $\frac{1}{\sqrt{2\pi}}y^{-\frac{1}{2}}e^{-\frac{y}{2}}.$
The dashed curves overlaying $P\left(\Delta G_{i}\right)$ in two
and three dimension are the prediction based on Eq. \ref{eq:P_DG_i}.}
\end{figure}

\section{Estimating the exponents}

To understand why the curves of $G/B$ versus $\Delta Z$ look like
approximate power laws for the four different pruning strategies,
we approximate Eq, \ref{eq:DMi} as a differential equation:

\begin{equation}
\frac{dM}{d\Delta Z}\approx-\alpha\frac{M}{\Delta Z},\label{eq:expon_eq}
\end{equation}
where $d\left(\Delta Z\right)=\frac{2}{N}$ is the change in the coordination
number when a bond is pruned and

\begin{equation}
\alpha\equiv\Delta Z\frac{N}{2}\frac{\tau_{i}^{2}}{S_{i}^{2}}.\label{eq:expon_form}
\end{equation}
Here $\tau_{i}^{2}=M_{i}/M$ is proportional to the energy on bond
$i$, but normalized so that $\sum_{i}\tau_{i}^{2}=1$. Typically,
the stresses due to a global deformation are not localized and therefore
$\tau_{i}^{2}\sim\frac{1}{N}$. We also note that~\cite{Hexner2017,WyartReview}
$S_{i}^{2}\propto\Delta Z$. For these two reasons, $\alpha$ should
not depend on the system size or $\Delta Z$. If $\alpha$ is constant
then the solution to this equation is $M\propto\Delta Z^{-\alpha}$.
If bonds are chosen with some specific rule and the distribution of
$\alpha$ is stationary, then its average remains constant as the
system is pruned. In this case, $M\propto\Delta Z^{-\overline{\alpha}}$,
where $\overline{\alpha}$ denotes the average of $\alpha$ and depends
on the pruning procedure.

In Ref. \cite{Hexner2017} the distributions of $\Delta B_{i}$ and
$\Delta G_{i}^{\left(j\right)}$ were measured for different pruning
procedures and it was shown that for different pruning strategies
the distribution of $\Delta G_{i}^{\left(j\right)}$ is universal
and is given by Eq. \ref{eq:dist_DG_i}. While the starting distribution
of $\Delta B_{i}$ is initially different, it evolves to this universal
distribution for all pruning strategies discussed here except for
the case when the bonds with $min\Delta B_{i}$ are targeted. Based
on the definition of $\alpha$ in Eq. \ref{eq:expon_form} $\alpha\propto\Delta M_{i}$,
and therefore $\alpha$ will have the same distribution 
\begin{equation}
P\left(\alpha\right)=\frac{1}{\sqrt{2\pi\alpha_{0}}}\alpha^{-1/2}exp\left(-\frac{\alpha}{2\alpha_{0}}\right).\label{eq:P_alpha}
\end{equation}
The only free parameter is $\alpha_{0}$ and it can be evaluated by
noting that if bonds are removed randomly, $\overline{\alpha}=1$
\cite{tuning_by_pruning}. By requiring that $\int d\alpha P\left(\alpha\right)\alpha=1$
we find that $\alpha_{0}=1$.

We begin by considering the exponents associated with pruning the
$min\Delta M_{i}$. Numerically, it is found that in all cases the
change $M$ is very small. Due to the weak correlation between $\Delta B_{i}$
and $\Delta G_{i}$, pruning the $min\Delta B_{i}$ ($min\Delta G_{i}$)
results in the decrease in $G$ ($B$) as if a random bond is removed.
Therefore, $G\propto\Delta Z^{\approx1.0}$ when the $min\Delta B_{i}$
bonds are removed and $B\propto\Delta Z^{\approx1.0}$ when the $min\Delta G_{i}$
bonds are removed.

The case of pruning $max\Delta M_{i}$ is very different. In general,
these bonds carry a lot of stress which is then redistributed upon
their removal. Ref. \cite{TunableFailure} shows that these stresses
are redistributed on the length scale of $\zeta$ which diverges in
the limit of $\Delta Z\rightarrow0$. Therefore, if the system size
is smaller than $\zeta$, then bonds are removed approximately homogeneously
throughout the system, while for systems larger than $\zeta$ a system
spanning crack forms. All our data and analysis concerns the first,
homogeneous case.

We begin by estimating the exponent associated with the change of
the bulk modulus $B\propto\Delta Z^{\overline{\alpha}_{max\Delta B_{i}}}$
for the $max\Delta B_{i}$ procedure. This requires a calculation
of the average maximal value of $\alpha$, which depends on the number
of independent bonds. Assuming $N$ independent random variables,
the distribution of $\alpha_{max}$ is given by:

\begin{equation}
Q\left(\alpha_{max}\right)=N\left(\int_{0}^{\alpha_{max}}dyP\left(y\right)\right)^{N-1}P\left(\alpha_{max}\right).\label{eq:max_dist}
\end{equation}
Its average can be estimated numerically, yielding $\overline{\alpha}_{max\Delta B_{i}}\approx14.6$
for $N=4000$ for comparison to simulations. This is somewhat greater
than the measured value of $11.5$ in two dimensions and $12.3$ in
three dimensions.

A similar estimate can be found for the case of pruning $max\Delta G_{i}$.
In two dimensions the distribution of $\alpha$, is then given by
$P\left(\alpha\right)=exp\left(-\alpha\right)$, where the average
is chosen to be unity. The distribution $Q\left(\alpha_{max}\right)$
can be computed analytically 
\begin{eqnarray}
Q\left(\alpha_{max}\right) & = & N\left(1-e^{-\alpha_{max}}\right)^{N-1}e^{-\alpha_{max}}\label{eq:Q_max}\\
 & \approx & Ne^{Ne^{-\alpha_{max}}}e^{-\alpha_{max}}
\end{eqnarray}
where the asymptotic form is the Gumbel distribution, with $\overline{\alpha}_{max}\approx logN+\gamma$,
and $\gamma\approx0.5772$ is the Euler\textendash Mascheroni constant.
This an instance of extreme value theory which predicts that when
$N$ is large, $Q\left(\alpha_{max}\right)$ will always be given
by a Gumbel distribution as long as $P\left(\alpha\right)$ decays
fast enough \cite{ExtremeValue}. In $2d$ the shear modulus should
therefore scale as $G\propto\Delta Z^{\overline{\alpha}_{max\Delta G_{i}}}$
with the exponent $\overline{\alpha}_{max\Delta G_{i}}\approx8.87$
(for $N=4000$ as used in simulations). This is the same order but
$36\%$ larger than the value $\approx6.5$ measured numerically.
A similar analysis in three dimensions yields $\overline{\alpha}_{max\Delta G_{i}}\approx5.0$
which is also $22\%$ greater than the value of $4.1$ found numerically.

Since in all cases the estimated exponents are larger than those found
numerically we test the validity of our assumptions. First, we consider
the distribution of $\alpha$ and check if it is indeed stationary,
focusing on the large $\alpha$ values crucial for the $max\Delta M_{i}$
pruning. To probe the tail of $P\left(\alpha\right)$ we measure $\overline{\alpha}_{max}$
as a function of bonds removed. This is found to be a constant, after
several bonds are removed.

Interestingly, the $\overline{\alpha}_{max}$ is found to be smaller
than that predicted from Eq. \ref{eq:max_dist}. We note that $\overline{\alpha}_{max}$
has a logarithmic dependence on the system size, $N$, since it samples
the exponential tail of $P\left(\alpha\right)$. Indeed, we find numerically
that the $\overline{\alpha}_{max\Delta B_{i}}$ grows slightly when
the system size is increased. However, contrary to our assumption
not all bonds are independent, and as demonstrated in Section \ref{sec:Spatial-correlations}
there are significant spatial correlations in the system. We believe
these correlation reduce $\overline{\alpha}_{max}$ , since they re-normalize
$N$ to a smaller value. Naively, if the spatial correlations are
the dominant contribution then the independent number of bonds should
scale as $N/\xi^{d}f_{\xi}$ where $\xi$ is the correlation length
associated with $\Delta B_{i}$ or $\Delta G_{i}$ and $f_{\xi}$
are the number of independent bonds within $\xi^{d}$.

\begin{figure}[H]
\includegraphics[scale=0.43]{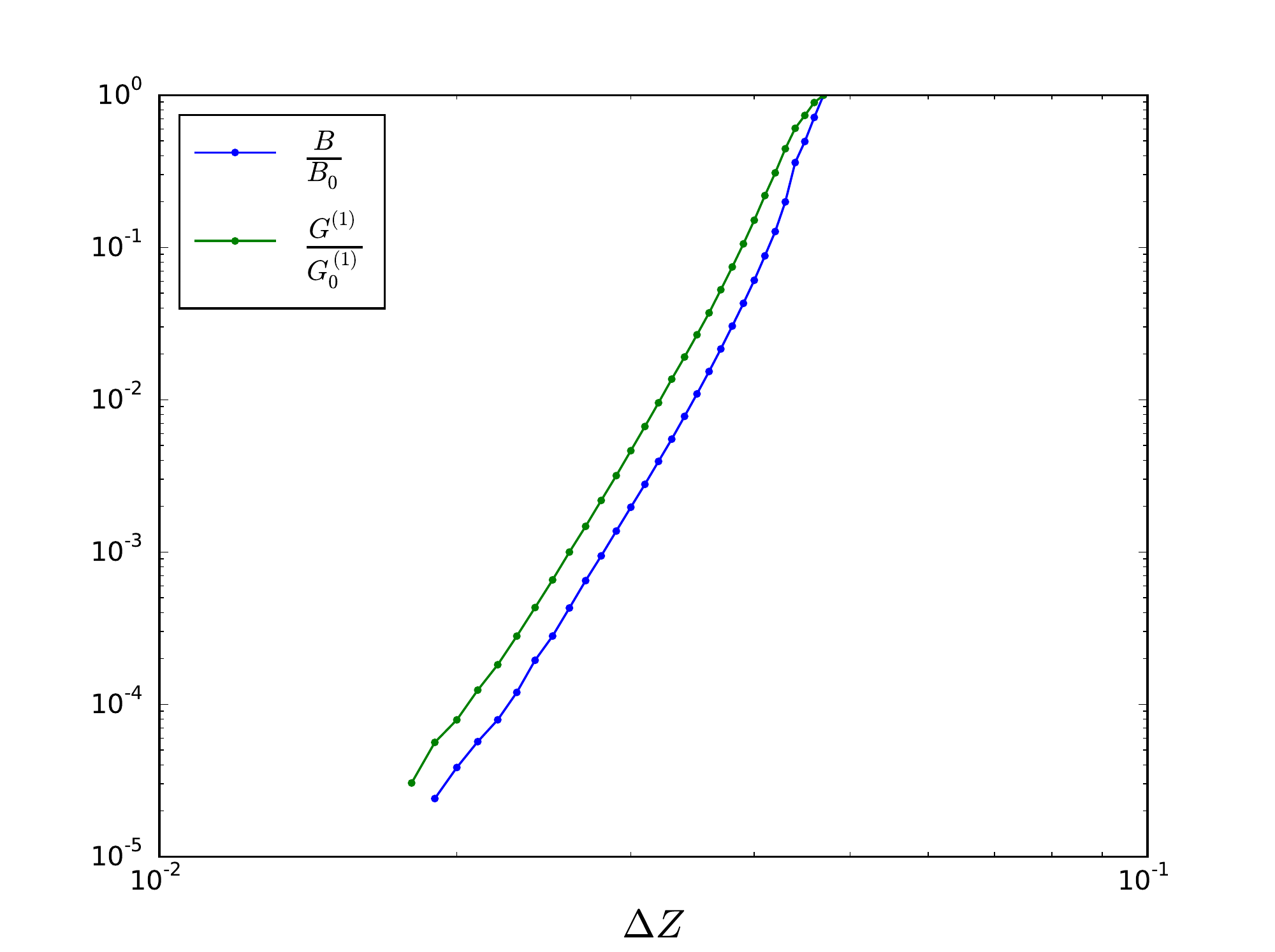}

\caption{A comparison between $B\left(\Delta Z\right)$ for the $max\Delta B_{i}$
pruning to the $G^{\left(1\right)}\left(\Delta Z\right)$ for the
$max\Delta G_{i}^{\left(1\right)}$ pruning. After several bonds are
removed these are almost parallel, suggesting universality. Here $G^{\left(1\right)}$
denotes the simple shear modulus and $B_{0}$ and $G_{0}^{\left(1\right)}$
denote the bulk and shear modulus before any pruning. \label{fig:BvsCxy}}
\end{figure}

We conclude by noting that despite the system size dependence there
are features that are universal, stemming from the $\alpha$ distribution
being independent of the pruning protocol. This is demonstrated by
comparing pruning $max\Delta B_{i}$ to $max\Delta G_{i}^{\left(1\right)}$,
where $G^{\left(1\right)}$ is the simple shear modulus. Since these
share the same $P\left(\alpha\right)$, we expect that $B\left(\Delta Z\right)$
should have the same behavior as $G^{\left(1\right)}\left(\Delta Z\right)$
after several bonds are removed. Indeed, in Fig. \ref{fig:BvsCxy}
these two curves shown to be almost parallel in the small $\Delta Z$
regime.

\section{Spatial correlations\label{sec:Spatial-correlations}}

In this section we discuss spatial correlations of $\Delta B_{i}$
and $\Delta G_{i}$. This is most easily done in Fourier space, for
which we define: 
\begin{equation}
\left\langle \left|\Delta M\left(q\right)\right|^{2}\right\rangle =\frac{1}{N\sigma_{\Delta M}^{2}}\left\langle \left|\sum_{i}\Delta M_{i}e^{-iqr}\right|^{2}\right\rangle \label{eq:corr_func_DBi_DGi}
\end{equation}
where $\sigma_{\Delta M}^{2}$ is the variance of $\Delta M_{i}$
which is a convenient normalization. The average is performed over
all directions in $q$ space and over about 10 realizations of disordered
unpruned networks. Aside from normalization, this is the Fourier transform
of the correlation function of $\left\langle \Delta M\left(r\right)\Delta M\left(0\right)\right\rangle -\left\langle \Delta M\right\rangle ^{2}$.
Since identifying growing length scales requires systems with a large
linear dimension, we focus on two-dimensional systems.

\begin{figure}[H]
\begin{centering}
\includegraphics[scale=0.45]{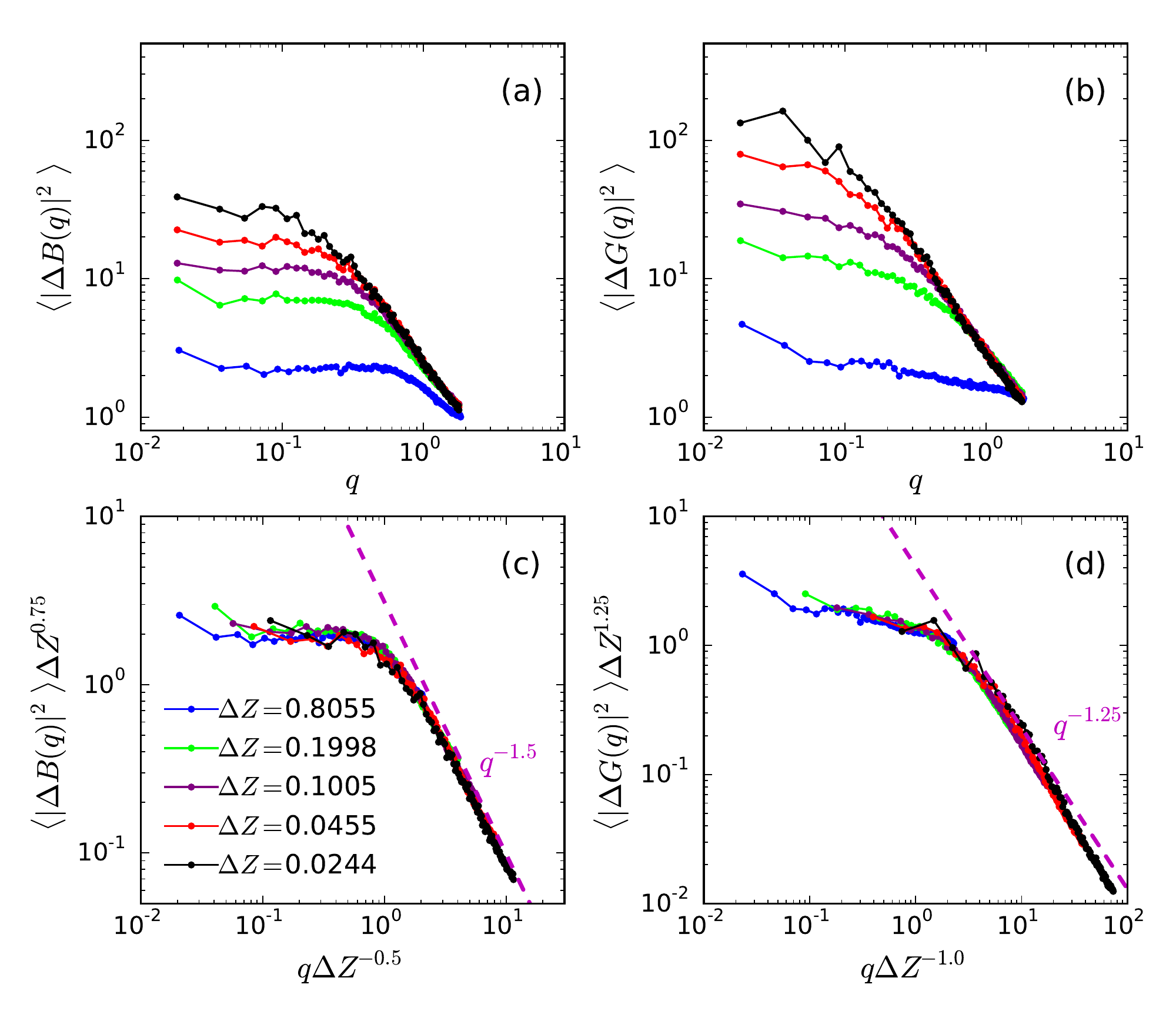} 
\par\end{centering}
\caption{The spatial correlations in two dimensions measured by the angle average
of $\Delta M\left(q\right)$ defined in Eq. \ref{eq:corr_func_DBi_DGi}.
In panel $(a)$ and $\left(b\right)$ we show the unscaled distributions
while in panel $\left(c\right)$ and $\left(d\right)$ we collapse
by rescaling the axis with powers of $\Delta Z$. The $\Delta B_{i}$
correlation length is consistent with $\xi_{\Delta B}\propto\Delta Z^{-0.5}$
while the $\Delta G_{i}$ correlation length is consistent with $\xi_{\Delta G}\propto\Delta Z^{-1.0}$.
\label{fig:SpatialCorrelations} }
\end{figure}

Figure \ref{fig:SpatialCorrelations} shows $\left\langle \left|\Delta B\left(q\right)\right|^{2}\right\rangle $
and $\left\langle \left|\Delta G\left(q\right)\right|^{2}\right\rangle $
in two dimensions for different values of $\Delta Z$. There are two
regimes. At large values of $q$ both quantities vary as $q^{-\gamma}$
with $\gamma_{\Delta B}\approx1.5$ for the bulk modulus and $\gamma_{\Delta G}\approx1.25$
for the shear modulus. This indicates spatial power-law correlations
which scale as, $r^{-d+\gamma}$ for $r$ smaller than a correlation
length, $\xi$. At small $q$ there is a crossover to a constant which
signals a transition to an uncorrelated state. The crossover value
of $q$ is identified as the inverse correlation length and its $\Delta Z$
dependence is found by a data collapse where the two axes are scaled
with a power of $\Delta Z$. We find that the $\Delta B_{i}$ correlation
length is given by $\xi_{\Delta B}\propto\Delta Z^{-0.5\pm0.15}$
while the $\Delta G_{i}$ correlation length scales as $\xi_{\Delta G}\propto\Delta Z^{-1.0\pm0.15}$.
This differs from the usual picture~\cite{Silbert2005}, in which
the bulk modulus is associated with the cutting length $\ell^{*}\propto\Delta Z^{-1}$
and the shear modulus with the transverse length scale $\ell^{\dagger}\propto\Delta Z^{-\frac{1}{2}}$.

\begin{figure}[H]
\begin{centering}
\includegraphics[scale=0.45]{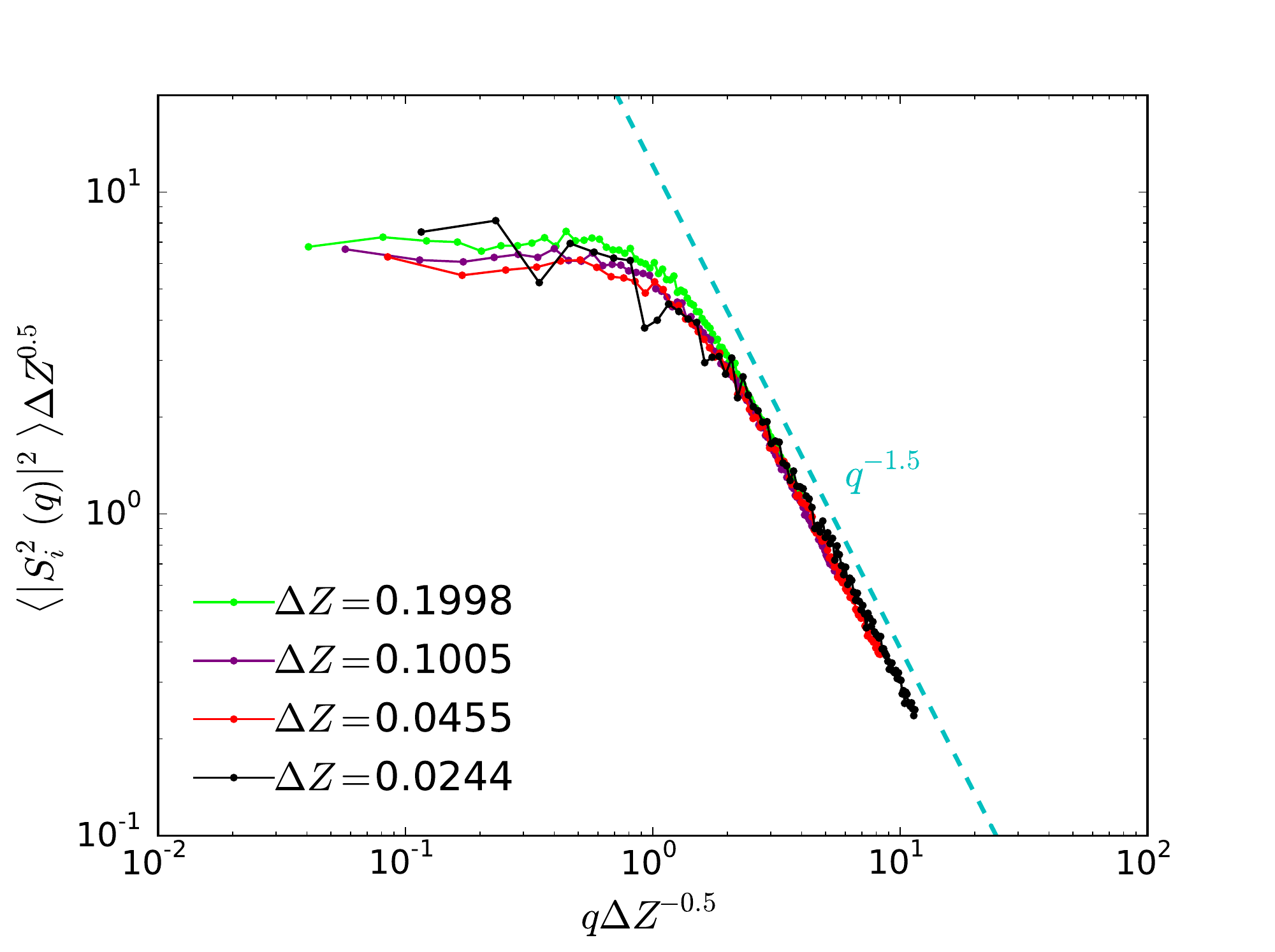} 
\par\end{centering}
\caption{The spatial correlations in two dimensions of $S_{i}^{2}$ measured
using $\left\langle \left|\Delta S_{i}^{2}\left(q\right)\right|^{2}\right\rangle $.
The correlation length agrees with $\Delta Z^{-\frac{1}{2}}$. \label{fig:S_ii_spatial_corr} }
\end{figure}

Since $\Delta M_{i}$ depends on $S_{i}^{2}$ we also measure the
correlations of $S_{i}^{2}$. Using the same definition in Eq. \ref{eq:corr_func_DBi_DGi}
we compute $\left\langle \left|\Delta S_{i}^{2}\left(q\right)\right|^{2}\right\rangle $
by replacing $\Delta M_{i}$ with $S_{i}^{2}$. These correlations
are shown in Fig. \ref{fig:S_ii_spatial_corr} and the diverging length
scale is consistent with $\Delta Z^{-0.5\pm0.15}$. Interestingly,
at large $q$, $\left\langle \left|\Delta S_{i}^{2}\left(q\right)\right|^{2}\right\rangle \propto q^{\approx-1.5}$
as in the case of $\left\langle \left|\Delta B\left(q\right)\right|^{2}\right\rangle $.
This suggests that these correlations have the same source. We also
note the length scale $\Delta Z^{-0.5}$ has been observed in the
tension profile resulting from squeezing a bond~\cite{dipole_response,Jamming_SSS}.
This should be the same length scale.

\section{Conclusions}

In this paper we have shown that pruning bonds with maximum or minimum
values of $\Delta B_{i}$ and $\Delta G_{i}$ provides an effective
method for designing disordered meta-materials with targeted properties.
Previous protocols~\cite{tuning_by_pruning} relied on targeting
the \textit{contributions} to the moduli, $B_{i}$ and $G_{i}$. The
current approach targets the \textit{change} in the moduli when a
bond $i$ is removed (i.e., targeting $\Delta B_{i}$ and $\Delta G_{i}$).
This procedure allows us to tune the network to a nearly maximally
negative Poisson's ratio by targeting bonds with very small $\Delta G_{i}$.
This was not possible with the previous approach and is a significant
improvement because it allows the network to be pruned to the auxetic
limit without developing cracks.

We have also shown that if the system is kept isotropic, the possibility
of removing a bond which contributes very little to the shear modulus
is greatly reduced. This problem is more severe in higher dimension,
yet even in three dimensions there are enough small $\Delta G_{i}$
bonds to yield an auxetic material for $min\Delta G_{i}$ pruning.

We have provided a rationale for why moduli tend to scale as power
laws with $\Delta Z$ when bonds are removed. We estimated the exponents
for different pruning strategies, and found agreement to within 30\%
of the measured values. Our analysis suggests that the exponents have
some universal features, however, they depend weakly on the system
size. We also argue that spatial correlations may reduce the exponents.

Finally we have examined the spatial correlations of $\Delta B_{i}$
and $\Delta G_{i}$ in two dimensions and identified diverging length
scales. The correlation length for $\Delta B_{i}$ is given by $\xi_{\Delta B}\propto\Delta Z^{\approx-0.5}$
while the correlation length for $\Delta G_{i}$ is given by $\xi_{\Delta G}\propto\Delta Z^{\approx-1.0}$.
This contrasts with the intuition~\cite{Silbert2005} that associates
the $\ell^{*}=\Delta Z^{-1}$ with the bulk modulus and $\ell^{\dagger}=\Delta Z^{-\frac{1}{2}}$
with the shear modulus. We also find that $S_{i}^{2}$ has a correlation
length of $\Delta Z^{-0.5}$. These spatial correlation reduce the
range of possible designs in the system. 
\begin{acknowledgments}
We thank C. P. Goodrich for important discussions. We acknowledge
support from the Simons Foundation for the collaboration ``Cracking
the Glass Problem'' award \#348125 (DH, SRN), the US Department of
Energy, Office of Basic Energy Sciences, Division of Materials Sciences
and Engineering under Award DE-FG02-05ER46199 (AJL), the Simons Foundation
\#327939 (AJL), and the University of Chicago MRSEC NSF DMR-1420709
(SRN). 
\end{acknowledgments}

\section{Appendix }

\subsection{An efficient algorithm for recomputing the elastic properties resulting
from bond removal }

In this section we provide an efficient algorithm for recomputing
the elastic response due to the removal of a bond. We employ the notation
of Ref. \cite{Hexner2017,Lubensky_rev}. The energy cost of a deformation
is given by:

\[
U=\frac{1}{2}k\sum_{i,j}e_{i}\left[S_{i}\right]_{j}e_{j}
\]
and the tension in a bond is given by: 
\begin{equation}
t_{i}=k\sum_{j=1}^{N}e_{j}\left[S_{i}\right]_{j}.\label{eq:tensions}
\end{equation}
Here $e_{i}$ is the affine extension of a bond and $\left[S_{i}\right]_{j}$
is the tension in bond $j$ resulting from a unit change in the equilibrium
length of bond $i$. All information regarding the elastic behavior
depends only on $\left[S_{i}\right]_{j}$ and therefore we would like
to compute how it changes with the removal of a bond. In particular
$S_{i}^{2}\equiv\left[S_{i}\right]_{i}$ and $M_{i}=\frac{2t_{i}^{2}}{V\epsilon^{2}}$
where $V$ is the volume and $\epsilon$ is the strain. Eq. 5 of Ref.
\cite{Hexner2017} allows to compute the change in the tension when
a bond is removed for any deformation. The corresponding affine extension
for this deformation is given by $e_{j}=\delta_{ij}$. Assuming bond
$k$ is removed the element $\left[S_{i}\right]_{j}'$ is then given
by:

\[
\left[S_{i}\right]_{j}'=\left[S_{i}\right]_{j}-\frac{\left[S_{i}\right]_{k}\left[S_{k}\right]_{j}}{\left[S_{k}\right]_{k}}.
\]
Thus the evolution of the whole elastic response is easily computed
when a bond is removed, including $B_{i}$, $G_{i}$ and $S_{i}^{2}$
.

\subsection{Distribution of the sum of random variables with the universal form\label{appendix_sum_rand_var}}

We compute the sum of $n$ independent random variables, $Z=\sum_{i=1}^{n}y_{i}$,
where $P\left(y_{i}\right)=\frac{1}{\sqrt{2\pi y_{0}}}y_{i}^{-1/2}exp\left(-\frac{y_{i}}{2y_{0}}\right)$.
We exploit, the fact that $x_{i}=\sqrt{y_{i}}$ is a Gaussian random
variable.

\begin{equation}
P\left(x_{i}\right)=\frac{1}{\sqrt{2\pi y_{0}}}exp\left(-\frac{x_{i}^{2}}{2y_{0}}\right).
\end{equation}
To compute the distribution of $Z$, we first compute the distribution
of $R=\sqrt{Z}=\sqrt{\sum x_{i}^{2}}$ and then use a transformation
of variables to compute $P\left(Z\right)$. The distribution of $R$
is the sum of Gaussian variables and therefore straightforward:

\begin{equation}
P\left(R\right)=\frac{S\left(n\right)}{\left(2\pi y_{0}\right)^{n/2}}R^{n-1}exp\left(-\frac{R^{2}}{2y_{0}}\right)
\end{equation}
where $S\left(n\right)=\frac{2\pi^{\frac{n}{2}}}{\Gamma\left(\frac{n}{2}\right)}$
is the surface of a n-dimensional hypersphere and $\Gamma$ is the
Gamma function. A transformation of variables results in:

\begin{align}
P\left(Z\right) & =P\left(R\right)\frac{dR}{dZ}=\frac{1}{2}\frac{S\left(n\right)}{\left(2\pi y_{0}\right)^{n/2}}Z^{n/2-1}exp-\frac{Z}{2y_{0}}.\\
 & =\frac{1}{\left(2y_{0}\right)^{n/2}\Gamma\left(\frac{n}{2}\right)}Z^{n/2-1}exp-\frac{Z}{2y_{0}}.
\end{align}

The average of $Z$ can be computed by noting that it is a sum of
$n$ identical random variables, and thus is given by $n\left\langle y_{i}\right\rangle =ny_{0}$.

 \bibliographystyle{apsrev4-1}
\bibliography{biblo}

\end{document}